\documentclass[]{spie}

\usepackage[]{graphicx,color}
\usepackage[]{amsmath}
\usepackage[]{amssymb}

\title{High contrast imaging at the LBT: the LEECH exoplanet imaging survey}

\author{
Andrew J. Skemer\authorinfo{\supit{*} Contact: askemer@as.arizona.edu}\supit{*a},
Philip Hinz\supit{a},
Simone Esposito\supit{b},
Michael F. Skrutskie\supit{c},
Denis Defrere\supit{a},
Vanessa Bailey\supit{a},
Jarron Leisenring\supit{a},
Daniel Apai\supit{a}, 
Beth Biller\supit{d,e}, 
Mickael Bonnefoy\supit{d,f}, 
Wolfgang Brandner\supit{d}, 
Esther Buenzli\supit{d},
Laird Close\supit{a},
Justin Crepp\supit{g},
Robert J. De Rosa\supit{h,i},
Silvano Desidera\supit{j},
Josh Eisner\supit{a},
Jonathan Fortney\supit{k},
Thomas Henning\supit{d},
Karl-Heinz Hofmann\supit{l},
Taisiya Kopytova\supit{d},
Anne-Lise Maire\supit{j},
Jared R. Males\supit{a},
Rafael Millan-Gabet\supit{m},
Katie Morzinski\supit{a},
Apurva Oza\supit{c},
Jenny Patience\supit{h},
Abhijith Rajan\supit{h},
George Rieke\supit{a},
Dieter Schertl\supit{l},
Joshua Schlieder\supit{d},
Kate Su\supit{a},
Amali Vaz\supit{a},
Kimberly Ward-Duong\supit{h},
Gerd Weigelt\supit{l},
Charles E. Woodward\supit{n}
and
Neil Zimmerman\supit{d,o}
\skiplinehalf
\supit{a} Steward Observatory, University of Arizona, Tucson, AZ, USA;
\skiplinehalf
\supit{b} Arcetri Observatory, Florence, Italy;
\skiplinehalf
\supit{c} University of Virginia, Charlotteville, VA, USA;
\skiplinehalf
\supit{d} Max Planck Institute for Astronomy, Heidelberg, Germany;
\skiplinehalf
\supit{e} University of Edinburgh, Edinburgh, UK;
\skiplinehalf
\supit{f} Institut de Planetologie et d'Astrophysique de Grenoble, Grenoble, France;
\skiplinehalf
\supit{g} Notre Dame University, South Bend, IN, USA;
\skiplinehalf
\supit{h} Arizona State University, Tempe, AZ, USA;
\skiplinehalf
\supit{i} University of Exeter, Exeter, UK;
\skiplinehalf
\supit{j} Padova Observatory, Padova, Italy;
\skiplinehalf
\supit{k} University of California, Santa Cruz, Santa Cruz, CA, USA;
\skiplinehalf
\supit{l} Max Planck Institute for Radio Astronomy, Bonn, Germany;
\skiplinehalf
\supit{m} NASA Exoplanet Science Institute, California Institute of Technology, Pasadena, CA, USA;
\skiplinehalf
\supit{n} Minnesota Institute for Astrophysics, University of Minnesota, Minneapolis, MN, USA;
\skiplinehalf
\supit{o} Princeton University, Princeton, NJ, USA;
}

\begin{document} 
\maketitle

\begin{abstract}
In Spring 2013, the LEECH (LBTI Exozodi Exoplanet Common Hunt) survey began its $\sim$130-night campaign from the Large Binocular Telescope (LBT) atop Mt Graham, Arizona. This survey benefits from the many technological achievements of the LBT, including two 8.4-meter mirrors on a single fixed mount, dual adaptive secondary mirrors for high Strehl performance, and a cold beam combiner to dramatically reduce the telescope's overall background emissivity. LEECH neatly complements other high-contrast planet imaging efforts by observing stars at L' (3.8 $\mu$m), as opposed to the shorter wavelength near-infrared bands (1-2.4 $\mu$m) of other surveys.  This portion of the spectrum offers deep mass sensitivity, especially around nearby adolescent ($\sim$0.1-1 Gyr) stars.  LEECH's contrast is competitive with other extreme adaptive optics systems, while providing an alternative survey strategy.  Additionally, LEECH is characterizing known exoplanetary systems with observations from 3-5$\mu$m in preparation for \textit{JWST}.
\end{abstract}

\section{INTRODUCTION\label{sec:intro}}
High contrast direct imaging offers the opportunity to study widely separated exoplanets in great detail.  The handful of systems that have been imaged\cite{2010Natur.468.1080M,2010Sci...329...57L,2013ApJ...774...11K,2013ApJ...779L..26R} now have photometry spanning the optical through thermal infrared\cite{2014ApJ...786...32M,2010ApJ...716..417H,2011ApJ...739L..41G} and low to medium resolution spectroscopy\cite{2011ApJ...733...65B,2013Sci...339.1398K}, all of which are contributing to our understanding of the appearance, composition and evolution of massive gas-giant planets.  Exoplanet characterization is still limited by the raw number and diversity of known systems.  We are now in a time where extreme adaptive optics systems and large surveys are expected to find new planets, pushing to cooler temperatures, lower masses and smaller separations.  There are several examples of dedicated planet finders (GPI\cite{2008SPIE.7015E..31M} , SPHERE\cite{2008SPIE.7014E..41B} , Project 1640\cite{2011PASP..123...74H}) and more general purpose facilities (LMIRcam\cite{2010SPIE.7735E.118S} , NACO\cite{2003SPIE.4839..140R} , HiCIAO\cite{2009AIPC.1158...11T}) well positioned to advance this field.  In this paper, we will describe the LBT Exozodi Exoplanet Common Hunt (LEECH), a large multinational survey to search for and characterize directly imaged planets in the ground-based mid-infrared (3-5$\mu$m).  

\section{INSTRUMENT AND OBSERVING STRATEGY\label{sec:instr}}
\begin{figure}[htbp]
\centering
      \includegraphics[width=0.80\linewidth]{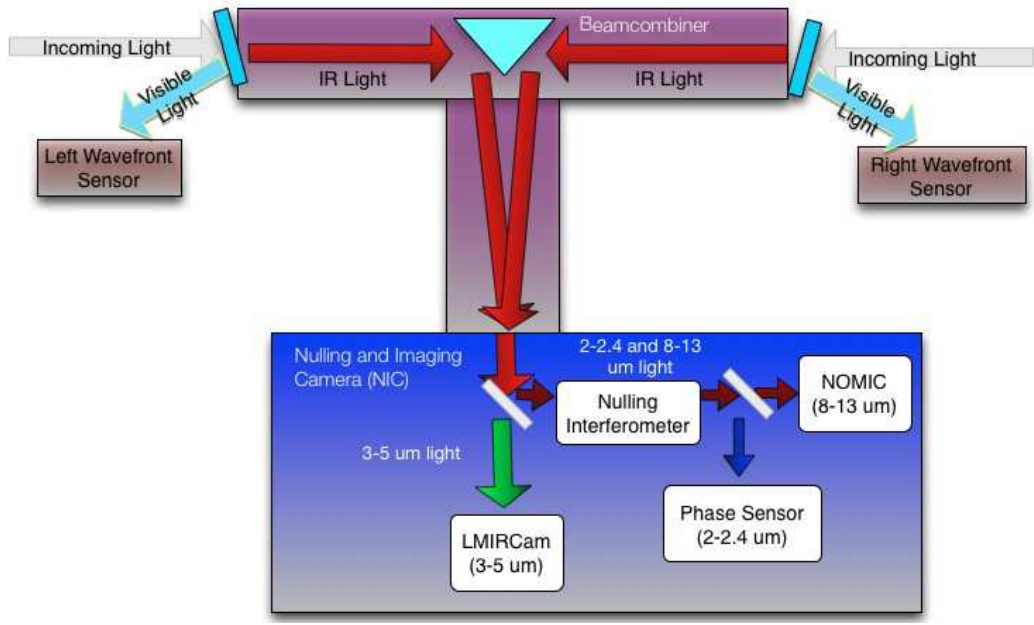}
      \caption{Block diagram of the Large Binocular Telescope Interferometer (LBTI) in the configuration used for LEECH (3-5$\mu$m light going to the 1-5$\mu$m capable LMIRcam science camera).  LBTI combines the light from the LBT's 2$\times$8.4 meter primary mirrors inside a large cryogenic structure that  maximizes sensitivity in a wavelength range that is usually dominated by telescope background emission.
	  } \label{fig:block}
\end{figure}

Of the major exoplanet imaging campaigns, LEECH is the only one operating at L' (3.8$\mu$m) where gas giant exoplanets peak in brightness and adaptive optics performance is superb.  At L', the telescope optics are significantly brighter than faint exoplanets, so special care must be taken to limit telescope emissivity and maximize sensitivity.  As with all exoplanet imaging surveys, contrast, the ability to detect a faint planet next to its bright host star, is paramount.  LEECH uses an assortment of subsystems to achieve the best L' (3.8$\mu$m) sensitivity and contrast in the World:
\begin{description}
  \item[The Large Binocular Telescope (LBT)] \hfill \\
    Located on Mount Graham, near Safford, Arizona, the Large Binocular Telescope features a pair of monolithic 8.4 meter primary mirrors on a common mount\cite{2014SPIE...VEILLET}.  Light from the two primaries can be directed to a single science instrument or two different science instruments.  When the two telescope beams are used together\cite{2014SPIE...HILL}, the LBT has a collecting area equivalent to a single 11.8 meter primary.  Coherently overlapping the two telescope beams further enhances sensitivity by making a diffraction pattern with 23 meter resolution in one axis, and 8.4 meter resolution in the orthogonal axis.
 \item[Twin Deformable Secondary Adaptive Optics Systems] \hfill \\
   Each primary mirror has a corresponding deformable secondary mirror, which can be held in a fixed position for seeing limited instruments, or driven with the LBT's adaptive optics system to correct atmospheric turbulence at 1 kHz.  Unlike most adaptive optics systems, deformable secondary AO systems are designed to have the minimum number of warm optics, attenuating thermal emission from the telescope to the point where adaptive optics is practical at wavelengths $>$2$\mu$m\cite{2000PASP..112..264L}.  With 672 actuators per side routinely correcting 400 modes, the LBTAO system is an extreme AO system capable of exceeding 80\% Strehl ratio at 1.6$\mu$m, 95\% Strehl ratio at 3.8$\mu$m, and 99\% Strehl ratio at 10$\mu$m\cite{2010ApOpt..49G.174E,2012SPIE.8447E..0UE}.  The LBTAO system feeds multiple science cameras, each of which has its own pyramid wavefront sensor. LBTI has two wavefront sensors: one for each side\cite{2014SPIE...BAILEY}.
 \item[Large Binocular Telescope Interferometer (LBTI)] \hfill \\
   Light bounces off of the LBT's primary mirrors and deformable secondary mirrors, and is then redirected by flat tertiary mirrors to the Large Binocular Telescope Interferometer (LBTI), which sits at a bent-Gregorian focus.  LBTI's entrance windows reflect optical light to the AO wavefront sensors and transmit infrared light into a cryogenic beam combiner, which brings the light from the two telescopes to focus and corrects path-length for interferometry without adding additional warm optics that would degrade sensitivity in the mid-infrared\cite{2014SPIE...HINZ}.  After passing through a large beam combiner (UBC; Universal Beam Combiner\cite{2008SPIE.7013E..67H}), the light enters a science instrument dewar (NIC; Nulling Infrared Camera\cite{2008SPIE.7013E.100H}), and then is directed to a 1-5 $\mu$m camera (LMIRcam; L/M-band Infrared Camera\cite{2012SPIE.8446E..4FL}) and/or a 10$\mu$m camera (NOMIC; Nulling Optimized Mid-Infrared Camera\cite{2014SPIE...HOFFMANN}) with the option of sending H or K-band light to a phase sensor\cite{2014SPIE...DEFRERE_PHASE}.  A block diagram of the system is shown in Figure \ref{fig:block}.  LBTI is designed to be a versatile instrument that can image the two telescope beams separately, overlap the two beams incoherently, overlap the beams coherently for wide-field (Fizeau) interferometry\cite{2014SPIE...LEISENRING}, or overlap the beams and pupils for nulling interferometry\cite{2014SPIE...HINZ}.  Due to the success of the LBTAO systems, LBTI is most commonly used as a general purpose adaptive optics imager, despite its name containing the word `Interferometer'.
 \item[L/M-band Infrared Camera (LMIRcam)] \hfill \\
    The science camera used by LEECH is LMIRcam, a general purpose 1-5$\mu$m imager/spectrograph with a fine plate scale (0.0107''/pixel) that Nyquist samples a K-band interferometric PSF and over-samples single aperture adaptive optics images\cite{2012SPIE.8446E..4FL}.  LMIRcam's detector is a 5.3$\mu$m cutoff Hawaii-2RG HgCdTe device.  The current detector electronics only readout 1024x1024 of the array's 2048x2048 pixels (an upgrade to SIDECAR-SAM electronics is planned).  LMIRcam has a variety of filters spanning 1-5$\mu$m, and a set of germanium grisms\cite{2012SPIE.8450E..3PK} that can produce R$\sim$400 spectra.  Additionally, LMIRcam has a set of Apodizing Phase Plate (APP) coronagraphs\cite{2010SPIE.7734E..85K}, and an Annular Groove Phase Mask (AGPM) coronagraph\cite{2014SPIE...DEFRERE_AGPM} that are in the testing phase and are not currently used for LEECH.
\end{description}

For most LEECH observations, we separate the two telescope beams onto the left and right halves of LMIRcam's detector, and nod up and down to remove the bright sky/telescope background (intermittent problems with the adaptive secondaries sometimes mean that we only use one side of the telescope\cite{2014SPIE...CHRISTOU}).  Each beam is reduced independently, as spatially variant static aberrations in the system need to be calibrated separately using angular differential imaging (ADI)\cite{2006ApJ...641..556M}.  The length of time in each nod is set to 30 seconds so that the $\sim$10 second nod overheads have only a minor impact on efficiency, and sky rotation is well-sampled in each beam for ADI.  Typical sequences last $\sim$2 hours to ensure adequate sky rotation for ADI, and depth to see faint planets against the bright L' background.  Most LEECH targets saturate the detector in its shortest efficient integration time ($\sim$0.3 seconds), so we bracket our observations with unsaturated images of the host stars taken with a 1\% transmission neutral density filter.  In cases where the star is still too bright, subframe readouts allow us to take images as short as 0.003 seconds, which when combined with a neutral density filter, will keep the brightest star on our target list (Sirius) within the detector's linear range.

\section{SYNERGY WITH HOSTS: Hunt for Observable Signatures of Terrestrial Systems\label{sec:HOSTS}}
\begin{figure}[htbp]
\centering
      \includegraphics[angle=90,width=0.6\linewidth]{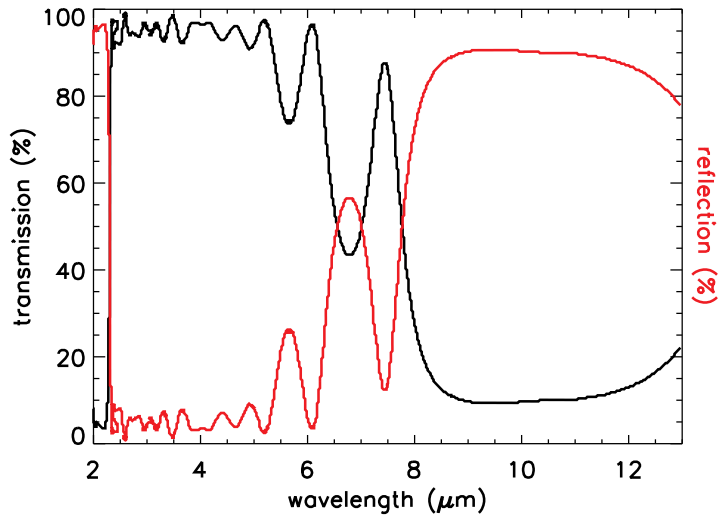}
      \caption{Transmission and reflection of LBTI's trichroic beam-splitter.  2$\mu$m light and 10$\mu$m light are reflected to the phase sensor and NOMIC science camera respectively, while 3-5$\mu$m light is transmitted to LMIRcam.  This configuration allows the LEECH exoplanet imaging survey to take data simultaneously with the HOSTS exozodi survey.
	  } \label{fig:trichroic}
\end{figure}

LBTI's other major project is HOSTS (the Hunt for Observable Signatures of Terrestrial Systems), a $\sim$60 night survey using LBTI's 10$\mu$m nulling interferometer to spatially resolve zodiacal emission from the habitable zones of nearby stars\cite{2014SPIE...DANCHI}.  HOSTS will survey the nearest stars to gauge the impact of zodiacal light on future space-based direct imaging missions.  LEECH was initially conceived as a ``piggy-back'' survey to the HOSTS survey.  The idea being that an innovative trichroic beam-splitter (see Figure \ref{fig:trichroic}) could reflect 2$\mu$m light to the fringe tracker and 10$\mu$m light to NOMIC, while transmitting 3-5$\mu$m light to LMIRcam, so that LEECH could observe simultaneously during HOSTS sequences.  In reality, direct-imaging with LBTI/LMIRcam and the LBTAO system was enticing enough to attract a collaboration of LBT partners to contribute $\sim$70 nights to the LEECH survey.  Thus the LEECH survey began in advance of the HOSTS survey, which is expected to begin in 2015.  When HOSTS begins, LEECH will execute ``piggy-back'' observations as planned.  The combination of nulling interferometry and direct imaging will allow studies of complete extrasolar systems, with inner-disk data coming from LBTI nulling, outer planet data coming from LBTI direct imaging, outer disk data coming from \textit{Spitzer}/\textit{Herschel}, and inner planet data coming from radial-velocity surveys.

\section{THE BENEFITS OF SEARCHING FOR EXOPLANETS AT L'\label{sec:Lp}}
\begin{figure}[htbp]
\centering
      \includegraphics[angle=90,width=0.69\linewidth]{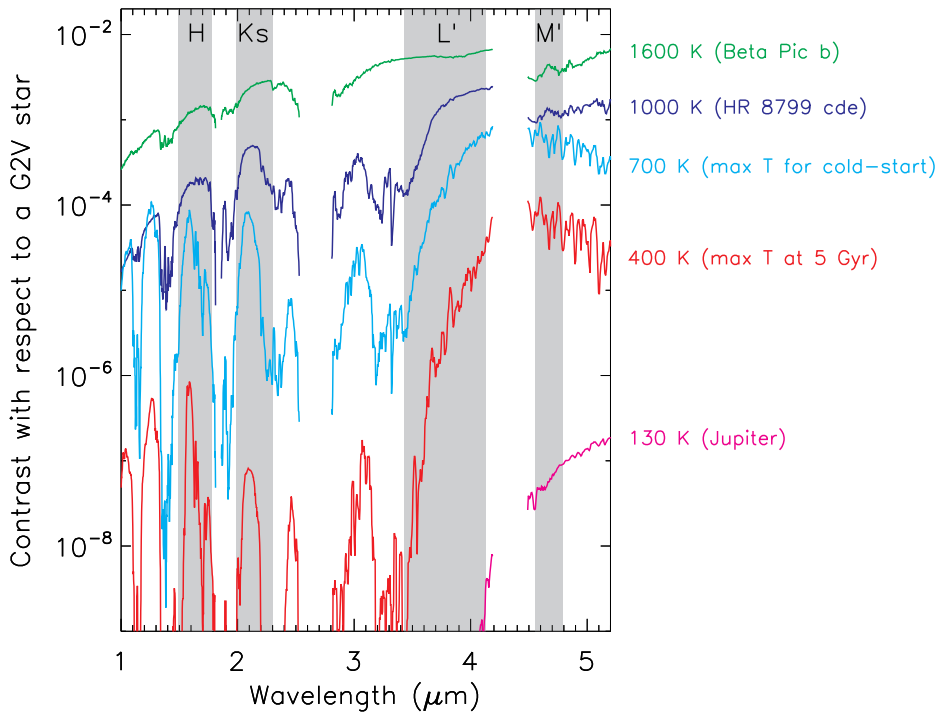}
      \vspace{0.25in}
      \caption{From Skemer et al. 2014\cite{2013arXiv1311.2085S}.  Characteristic examples of exoplanet-to-star contrasts (i.e. flux ratios) as a function of wavelength\cite{2011ApJ...737...34M,2003ApJ...596..587B}, showing (1) that gas-giant exoplanets can be detected with lower contrasts in the mid-infrared (3-5$\mu$m) than in the near-infrared (1-2$\mu$m), and (2) that this difference increases at lower temperatures.  While the planets that have been directly imaged to date ($\beta$ Pic b and HR 8799 c, d and e on this plot) are relatively warm (1600 K and 1000 K, respectively), it is likely that the majority of self-luminous exoplanets are much cooler.  Planets that formed by core-accretion (approximated by the ``cold-start'' models\cite{2008ApJ...683.1104F}) are never hotter than $\sim$700 K.  Planets around average-aged stars (5 Gyr) are never hotter than $\sim$400 K, regardless of formation history.  Jupiter, which may be a ubiquitous outcome of planet formation, is only $\sim$130 K. 
	  } \label{fig:L'just}
\end{figure}

Because gas-giant planets are cooler than their host stars, their contrast (brightness relative to their host stars) increases with wavelength until the Raleigh-Jeans tail, modulo opacity effects.  The 3-5$\mu$m window is the longest wavelength range accessible to ground-based high-contrast imagers, and also corresponds to a bright region for gas-giant exoplanets.  Figure \ref{fig:L'just} shows some characteristic examples of exoplanet contrasts as a function of wavelength.  The hottest examples correspond to some of the currently known directly imaged planets, which are known to be rare at wide separations.  Planets that form through core-accretion (approximated on this figure as ``cold-start'' models\cite{2008ApJ...683.1104F}) and planets around the vast majority of stars that are older than $\sim$1 Gyr will all be cooler than the current class of directly-imaged planets\cite{1998AA...337..403B}.  For all of the planets shown in Figure \ref{fig:L'just}, exoplanet contrasts at LEECH's operating wavelength (3.8$\mu$m) are lower than at 1.6$\mu$m, where most other direct-imaging surveys operate.  This property increases towards lower temperatures, where exoplanets are likely to be much more common.  The benefits of working at 3.8$\mu$m vs. 1.6$\mu$m for the LBTAO system are discussed explicitly in Section 5.

\section{PERFORMANCE\label{sec:performance}}

\begin{figure}[htbp]
\centering
      \includegraphics[angle=90,width=0.48\linewidth]{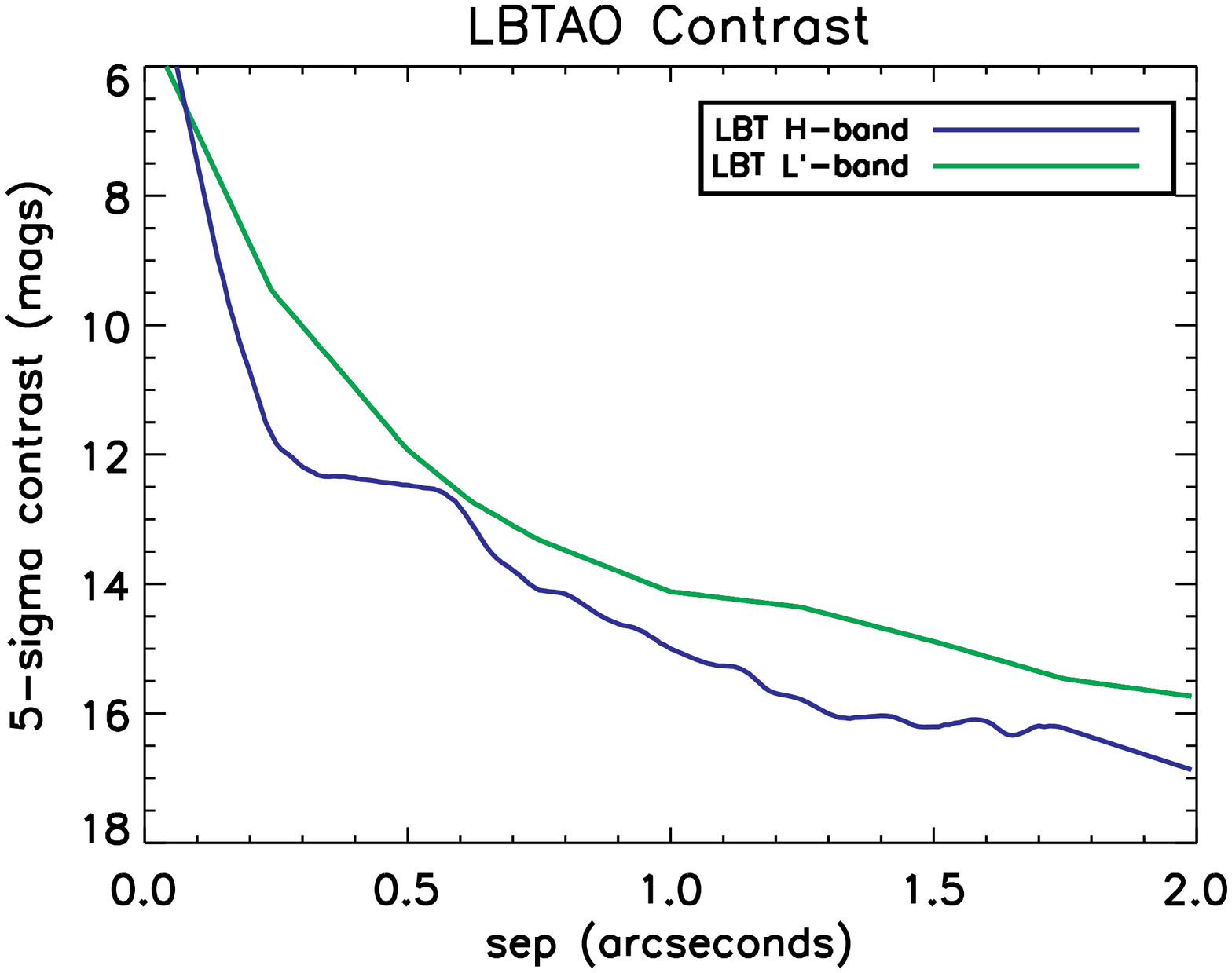}
      \includegraphics[angle=90,width=0.48\linewidth]{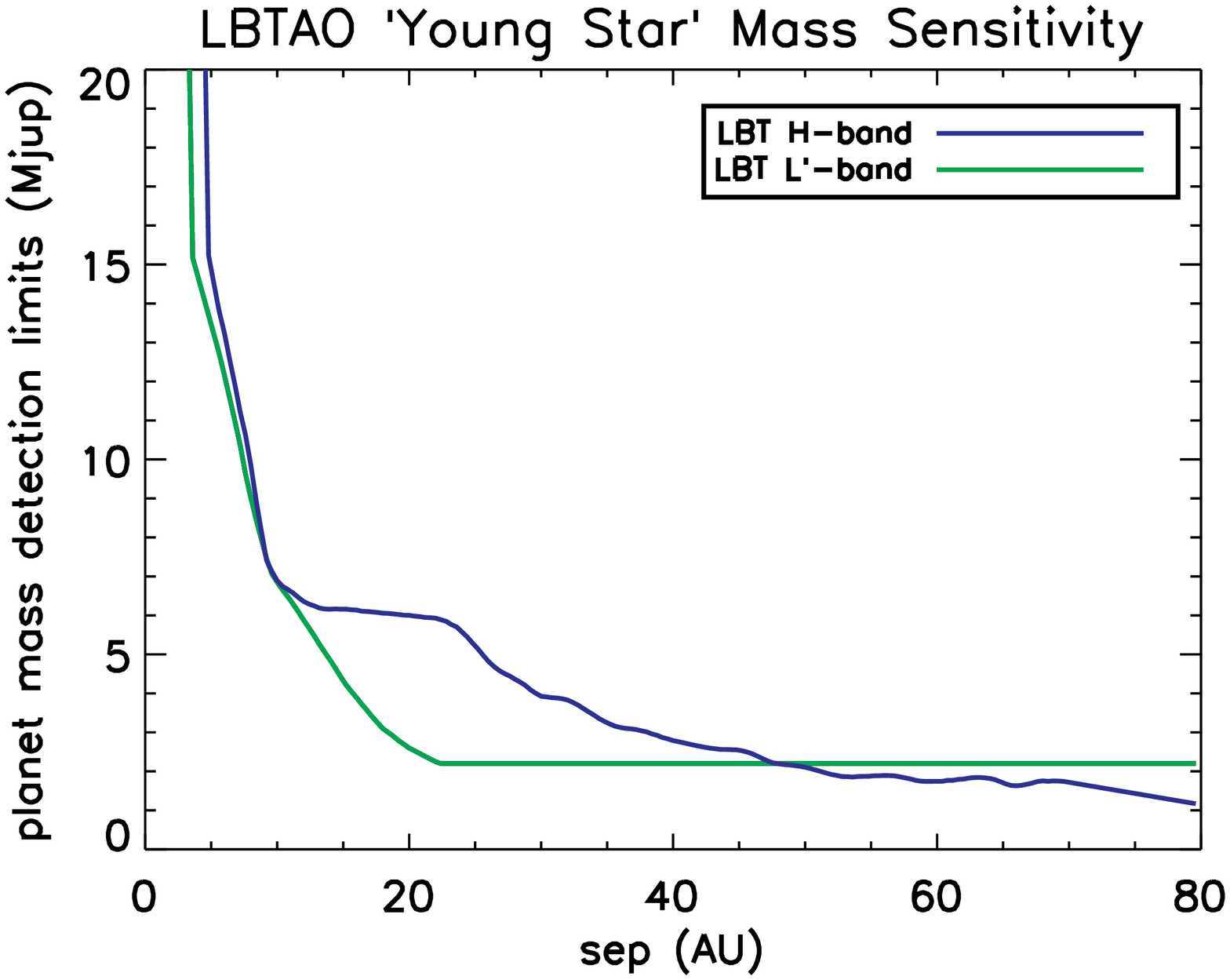}
      \includegraphics[angle=90,width=0.48\linewidth]{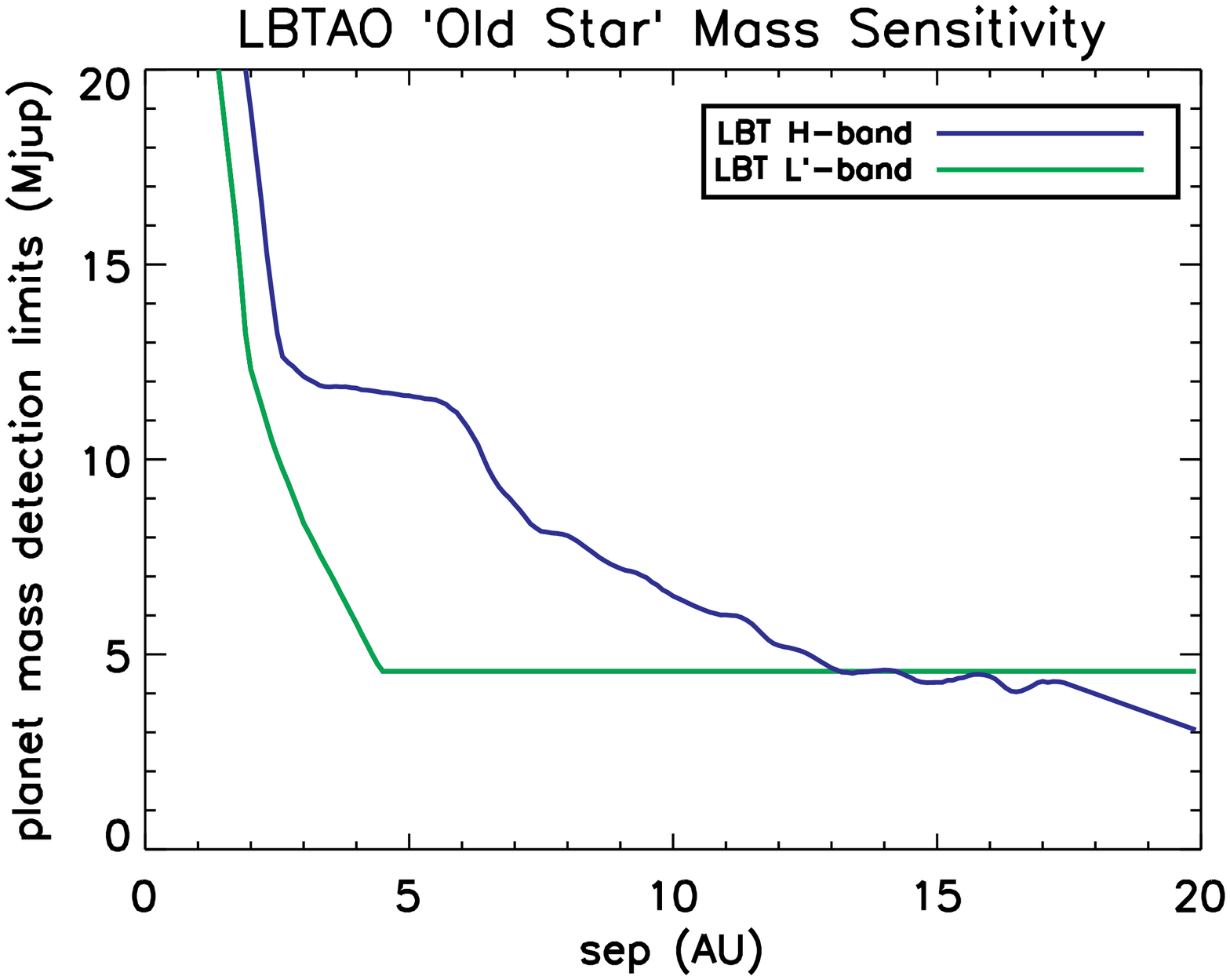}
      \includegraphics[angle=90,width=0.48\linewidth]{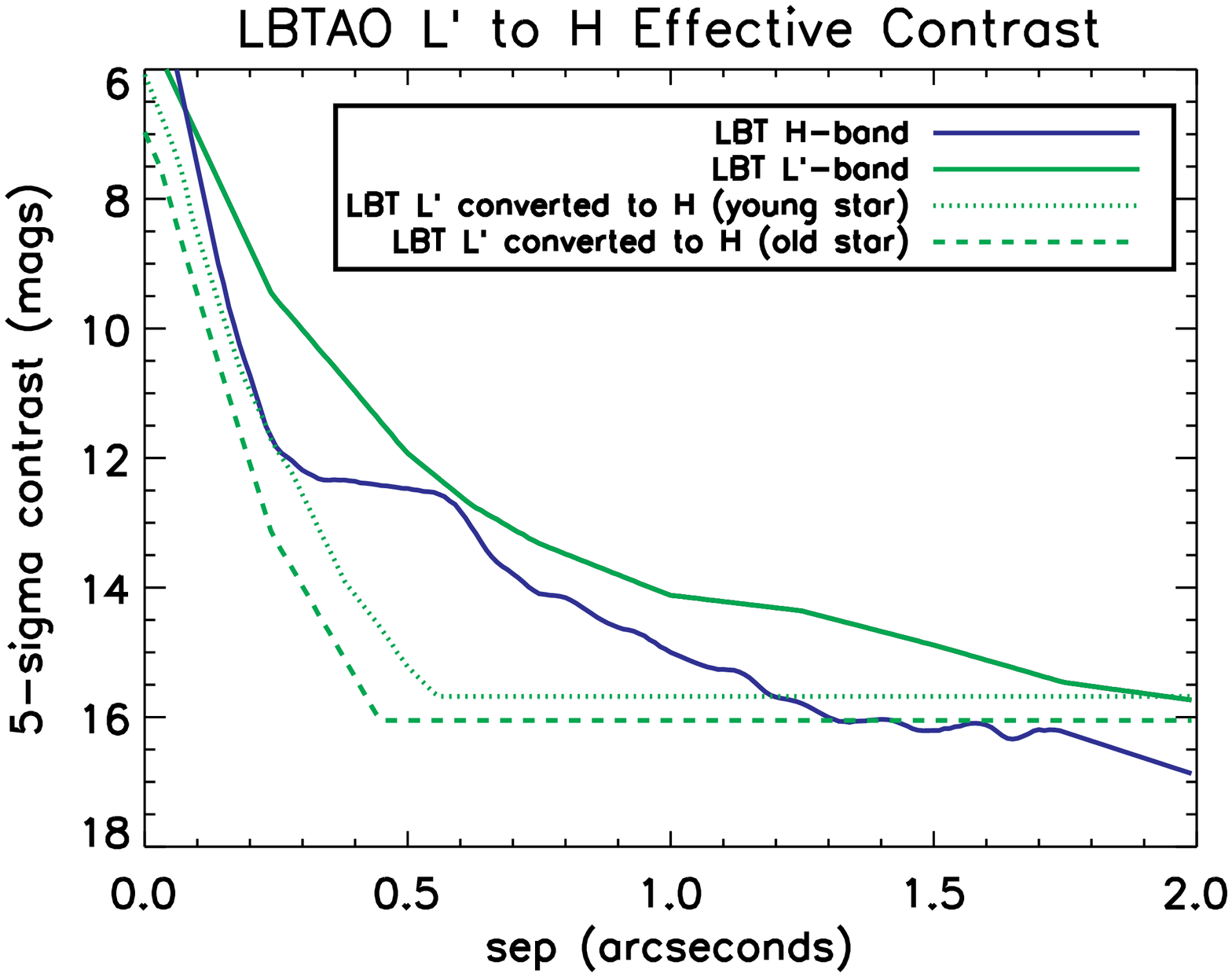}
      \caption{
        Top Left: On-sky contrast curves for LBTAO at H-band and L' (1.6$\mu$m and 3.8$\mu$m).  
        Top Right: Contrast curves converted to mass sensitivity curves for a hypothetical young star (30 Myr A-star with ``hot-start''\cite{1998AA...337..403B} planets at 40 pc, which is meant to approximate HR 8799).  Because a planet's contrast with respect to its host star decreases at longer wavelengths, the L' mass sensitivity is superior, even though the raw contrast curves show a deeper contrast at H-band.  Note that the floor on the contrast curve results from sky/telescope background noise, not quasi-static speckle residuals.
        Bottom Left: Contrast curves converted to mass sensitivity curves for a hypothetical old star (1 Gyr G-star with ``hot-start'' planets at 10 pc).  Even for old planets, LEECH achieves an excellent planet sensitivity.  And because old stars can be found near the Sun, the physical resolution is improved compared to the more distant young-star case.
        Bottom Right: Same as top-left, but with additional curves showing the L' contrast curve converted into an effective H-band contrast curve, based on the ``young star'' and ``old star'' examples shown in the other panels.  These effective H-band curves show the equivalent H-band contrast necessary to detect planets that are detectable in our L' data.  LEECH's effective H-band contrast is competitive with the best H-band contrasts achieved to date.
	  } \label{fig:contrast}
\end{figure}

Contrast curves for the LBTAO system are shown at H-band and L' in Figure \ref{fig:contrast}.  The H-band contrast curve\cite{2012ApJ...753...14S} uses the PISCES infrared camera with the FLAO/LUCI wavefront sensors.  The L' contrast curve uses LBTI and the LBTI wavefront sensors, which are copies of the FLAO wavefront sensors.  At L', the bright sky/telescope background precludes making a contrast curve at wide separations for all but the brightest stars.  As a result the L' contrast curve is necessarily based on data of Vega, our brightest star observed to date.  Unfortunately, the Vega data suffered from some instrumental artifacts that degraded our contrast.  So the shape of the curve uses the Vega data but the vertical position of the contrast curve is based on median contrasts from other observations taken in good conditions.  Note that for LEECH, the brightness of the target does not affect our contrast at small separations (the pyramid wavefront sensors produce the same performance from R$\sim$0-7 magnitudes), but for faint stars, the sky background becomes an important factor at large separations.

The H-band contrast, in magnitudes, exceeds the L' contrast curve at all separations.  However, when the magnitude contrasts are converted to mass contrasts using models\cite{1998AA...337..403B} similar to those shown in Figure \ref{fig:L'just}, the L' contrast exceeds the H-band contrasts at all separations until (for most stars) the bright sky background overtakes the residual speckle noise (see Figure \ref{fig:contrast}).  The L' curve can also be converted into an effective H-band contrast curve by determining what mass/temperature planet is detectable at L' and what corresponding H-band contrast would be necessary to detect that planet (see Figure \ref{fig:contrast}).  For separations where the LBT's L' effective H-band contrast exceeds its true H-band contrast, L' observations are superior.  At other separations, H-band observations are superior.  As shown in Figure \ref{fig:contrast}, the L' observations are superior at small separations (where gas-giant planets are likely to be more common), and H-band observations are superior at wide separations, where the L' observations are hampered by the bright sky/telescope background.

\begin{figure}[htbp]
\centering
      \includegraphics[width=1.0\linewidth]{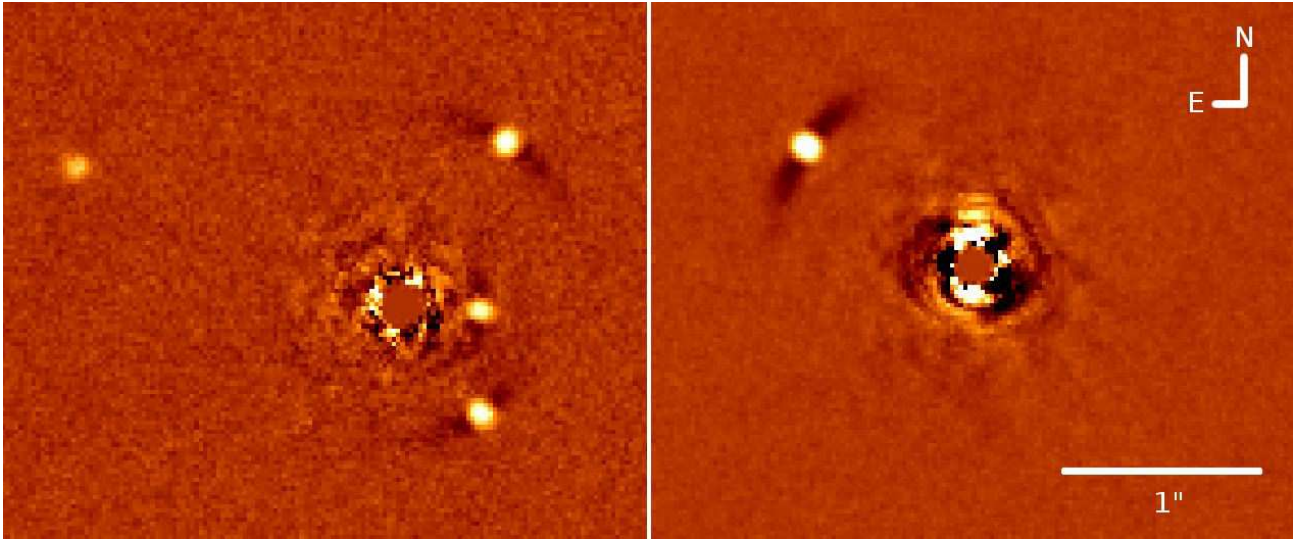}
      \caption{LEECH L' (3.8$\mu$m) images of the planetary systems around HR 8799 and Kappa And.  These images are the highest S/N images of either system, and show the best sensitivity to unseen inner planets for either system.
	  } \label{fig:8799}
\end{figure}

LEECH's effective H-band contrast is $\sim$16 magnitudes at 0.5'' for the examples shown, which is competitive with the contrasts currently being achieved by GPI\footnote{Early GPI commissioning data have a contrast of $\Delta$H$\sim$13.5 mags at 0.5'' for a flat spectrum and $\Delta$H$\sim$14.5 mags for a fully methanated spectrum in 30 minutes clock time\cite{2014arXiv1403.7520M}, which will be a typical exposure for the GPI survey. LBTI/LMIRcam's effective contrast of $\Delta$H$\sim$16 assumes a 2 hour integration, which is typical for LEECH.} and SPHERE.  Observations of known exoplanetary systems are consistently the highest S/N images taken of those systems (see, for example, Figure \ref{fig:8799}).

\section{SURVEY DESIGN AND TARGET LIST\label{sec:target}}
Previous surveys (NICI\cite{2013ApJ...777..160B,2013ApJ...776....4N,2013ApJ...773..179W} , IDPS\cite{2012A&A...544A...9V}, etc.) were designed to maximize the detections of exoplanets assuming an optimistic model for the evolution of planet luminosities (hot-start\cite{1998AA...337..403B}).  In the hot-start model, planets around young stars are very bright, and then rapidly fade as they evolve.  A more pessimistic model (cold-start\cite{2008ApJ...683.1104F}) predicts that planets start fainter than the hot-start models, but then fade more slowly.  Designing a survey to maximize hot-start detections pushes a target list towards young systems, even if they are far away, and angularly small on the sky.  However, the previous surveys yielded mostly non-detections, which could either mean that wide separation massive planets are extremely rare, or that wide separation massive planets don't evolve following the hot-start models.  To distinguish between these hypotheses, LEECH is designed to detect cold-start planets, so that non-detections rule out the presence of planets regardless of formation scenario.

By working at L', LEECH is optimized for detecting cold, low-luminosity exoplanets (see Figure \ref{fig:L'just}).  To detect cold-start planets LEECH must have the sensitivity to detect 500-600 K planets, which corresponds to an absolute magnitude limit of L'$\sim$15 magnitudes\cite{1998AA...337..403B}.  The bright sky/telescope background limits typical LEECH observations to a sensitivity of L'$\sim$17.5 magnitudes, which implies a distance cut of 30 pc.  In reality, planets may follow evolutionary models that are intermediate between hot-start and cold-start models (e.g. warm-start models\cite{2012ApJ...745..174S}), so our target list is also designed to survey likely planet hosts (such as massive stars) out to larger distances.

The LEECH survey has a fluid target list consisting of nearby, intermediate-aged stars complementing other high-contrast imaging surveys of more distant, young stars. Some representative examples of our current target list include 52 members of the 300-600Myr Ursa Majoris moving group\cite{1993ApJ...402L...5S,2003AJ....125.1980K,2004ARAA..42..685Z}, 54 A and B-type stars within 55pc, and 30 young ($\le$1Gyr) FGK stars within 25pc. The Ursa Majoris (UMa) moving group targets were drawn from a membership list\cite{2003AJ....125.1980K}, and span a range of spectral types from A0 to M2. The sample is biased towards more massive stars, with 44\% of the sample being earlier than spectral type A9, a higher proportion than the canonical initial mass function\cite{2001MNRAS.322..231K}. The common age for the UMa members strengthens the statistical analysis of the results by eliminating the range of companion mass limits assigned to a given contrast detection limit. The sample of nearby A and B-type stars was drawn from the union of the Tycho2\cite{2000AA...355L..27H} and \textit{Hipparcos}\cite{2007AA...474..653V} catalogs.  To ensure the sample included the brightest nearby stars, the first supplement to the Tycho2 catalog was also queried. An initial sample was constructed by selecting only those stars with a color consistent with a spectral type of A9 or earlier ({\it B}$-${\it V}$<$0.33\cite{1992oasp.book.....G}), visible from the Northern Hemisphere ($\delta \ge$-20degrees). For the A-type stars within the sample ({\it B}$-${\it V}$\geq$0), a distance limit of $D < $30pc was applied. For the B-type stars within the sample ({\it B}$-${\it V}$<0$), a larger distance limit of $D < $55$\sim$pc was applied, due to their relatively low frequency within the solar neighborhood. Stars with a larger distance uncertainty ($\sigma_{\pi}/\pi \ge 5\%$) were removed, along with two nearby white dwarfs. The nearby A-type stars Sirius and Vega were manually added to the sample, as these stars did not have entries within either the main Tycho2 catalogue, or its supplements. A final age criterion was applied to the sample members based on their position on the color-magnitude diagram, with those older than 1Gyr removed.  Additional young, nearby FGK stars were also included\cite{2008ApJ...687.1264M,2010ApJ...714.1551H}.

\section{CHARACTERIZATION\label{sec:characterization}}
In addition to searching for new exoplanets, LEECH is characterizing known directly imaged planets from 3-5$\mu$m, where LBTI/LMIRcam is and will continue to be the most powerful high-contrast imager until \textit{JWST}.  As shown in Figure \ref{fig:L'just}, gas-giant exoplanets emit the majority of their light from 3-5$\mu$m, making photometry at these wavelengths critical for measuring the bolometric luminosities of self-luminous exoplanets.  Early studies of directly-imaged planets show that they can look quite different from similar temperature field brown dwarfs, particularly at 3.3$\mu$m, where dusty models that include equilibrium methane absorption under-predict HR 8799 bcde and 2M1207 b's fluxes by more than a factor of 2\cite{2012ApJ...753...14S,2013arXiv1311.2085S}.  What we are learning about the handful of directly imaged planets now will be critical for planning observations and advancing our theoretical understanding of self-luminous planets in advance of \textit{JWST}.

\section{SUMMARY\label{sec:summary}}
LEECH is a $\sim$130 night, multinational survey to search for directly-imaged exoplanets using the Large Binocular Telescope, its 2 deformable secondary AO systems, the Large Binocular Telescope Interferometer, and its 1-5$\mu$m science camera, LMIRcam.  LEECH is the only major survey currently operating at 3.8$\mu$m, where gas-giant exoplanets peak in brightness and adaptive optics performance is superb.  Indeed, a comparison of the LBT's H-band and L'-band contrast curves confirms that for LBTAO, L' provides better mass contrasts than H-band, implying that it is the preferred wavelength for direct-imaging surveys.  Converting the L' contrast curve into an effective H-band contrast curve shows that we are as sensitive as an instrument providing $\Delta$H$\sim$16 magnitudes at 0.5'', which is competitive with other direct imaging surveys such as GPI and SPHERE.  The LEECH survey has been active since early 2013, characterizing known exoplanetary systems and searching for new ones.

\section*{ACKNOWLEDGMENTS}
The authors thank Bruce Macintosh for his helpful comments.  The LBT is an international collaboration among institutions in the United States, Italy and Germany. LBT Corporation partners are: The University of Arizona on behalf of the Arizona university system; Istituto Nazionale di Astrosica, Italy; LBT Beteiligungsgesellschaft, Germany, representing the Max-Planck Society, the Astrophysical Institute Potsdam, and Heidelberg University; The Ohio State University, and The Research Corporation, on behalf of The University of Notre Dame, University of Minnesota and University of Virginia.  This research was supported by NASA's \textit{Origins of Solar Systems} Program, grant NNX13AJ17G.  The Large Binocular Telescope Interferometer is funded by the National Aeronautics and Space Administration as part of its Exoplanet Exploration program.  LMIRcam is funded by the National Science Foundation through grant NSF AST-0705296.  EB is supported by the Swiss National Science Foundation (SNSF).

\bibliography{database}
\bibliographystyle{spiebib}

\end{document}